\documentstyle[preprint,aps]{revtex}

\catcode`\@=11
\def\numberbysection{\@addtoreset{equation}{section}
        \def\theequation{\thesection.\arabic{equation}}}
\def\beq{\begin{equation}}
\def\eeq{\end{equation}}
\def\barr{\begin{eqnarray}}
\def\earr{\end{eqnarray}}
\def\bea{\begin{eqnarray}}
\def\ena{\end{eqnarray}}

\def\u1{\widehat{U(1)}}
\def\su2{\widehat{SU(2)}_1}

\def\w3{{\cal W}_3}

\begin{document}
\draft
\title{Effective Chern-Simons Theories of Pfaffian and Parafermionic Quantum Hall States, 
and Orbifold Conformal Field Theories}
\author{Eduardo Fradkin$^1$, 
Marina Huerta$^{2,3}$\footnote{Present address: Dept.\ of Physics, 
Theoretical Physics, 
Oxford University, 1 Keble Road, Oxford, OX1 3NP, U.\ K.\ }, and Guillermo~R.~Zemba$^{4,5}$}
\address{$^1$ Dept.\ of Physics, University of Illinois, 1110 W.\ Green St.\ ,
Urbana, IL 61801-3080, U.S.A; 
$^2$Centro At\'omico Bariloche-C.\ N.\ E.\ A.\
and $^3$Instituto Balseiro, Universidad Nacional de Cuyo,
8400 S.\ C.\ de Bariloche, R\'{\i}o Negro, Argentina;
$^4$Depto.\ de F{\'\i}sica-C.\ N.\ E.\ A.\ , Av.\ Libertador 8250, 
1429 Buenos Aires, and
$^5$Universidad Favaloro, Sol{\'{\i}}s 453, 1078 Buenos Aires, Argentina}
\date{\today}

\maketitle

\begin{abstract}
We present a pure Chern-Simons formulation of families of
interesting Conformal Field Theories describing edge
states of non-Abelian Quantum Hall states. These theories contain
two Abelian Chern-Simons fields describing the
electromagnetically charged and neutral sectors of these
models, respectively. The charged sector is the usual Abelian Chern-Simons
theory that successfully describes Laughlin-type incompressible
fluids. The neutral sector is a $2+1$-dimensional theory
analogous to the $1+1$-dimensional orbifold conformal field
theories. It is based on the gauge group
$O(2)$ which contains a ${\bf Z}_2$ disconnected group
manifold, which is the salient feature of this theory.
At level $q$, the Abelian theory of the neutral sector gives 
rise to a ${\bf Z}_{2q}$
symmetry, which is further reduced by imposing the ${\bf
Z}_2$ symmetry of charge-conjugation invariance. The
remaining ${\bf Z}_q$ symmetry of the neutral sector is
the origin of the non-Abelian statistics
of the (fermionic) $q$-Pfaffian states.
\end{abstract}
\vfill
\narrowtext
%
\section{Introduction}

Non-Abelian statistics is an intriguing concept that perhaps may be
realized experimentally in two-dimensional electron gases (2DEG) in large
magnetic
fields. The leading candidates are the fractional quantum Hall (FQH)
states at filling factor $\nu=5/2$ and a number of other unusual
cases \cite{NAFQHE}.
A series of quantum Hall states displaying non-Abelian
statistics have been proposed for these states, the best known and
simplest of them being the Pfaffian FQH state \cite{more,grwi}.
However, still much about the nature of these states remains to be understood
in greater detail. All of the wave functions of the proposed states
show some sort of pairing correlation, as well as multiparticle generalizations
as in the case of the parafermionic states \cite{rere}. In all of these cases
the
spectrum of low-lying states contain excitations with non-Abelian
statistics. The notion of non-Abelian statistics \cite{more} is exhibited in
these FQH states
by exploiting an analogy between the wave functions of interacting
electrons in the lowest Landau level and correlators
in conformal field theory (CFT) \cite{cft}.

Although much work has been done on this problem, a general
understanding of what particular correlations are responsible for
non-Abelian statistics is still lacking. In particular, it is not 
known as
to what extent pairing-type mechanisms and its generalizations are
needed for non-Abelian statistics to exist. Specifically, more
than one physical realization of non-Abelian
statistics may be possible and several of them have been
proposed \cite{grwi,nawi,ctz5,wen-projection,cspf2}.
It is, therefore, an important issue to understand how these 
contructions are
related to each other and as to what extent they are truly different.

On the one hand, much of our present understanding of non-Abelian 
FQH states comes from the CFT structure of their edge states, 
which has been worked out in a number of
explicit examples \cite{more,rere,nawi,mire}.
It has been known from extensive work in the Abelian FQH states
\cite{wen} that CFT describes successfully the edge excitations, 
the low-energy degrees of freedom of FQH states.

On the other hand, it has been proven very useful to consider
a description of quantum Hall states in terms of an
effective low energy field theory describing the basic bulk
quasi-particle excitations \cite{wen,zhk,frohlich,lofra}.
These effective field theories generically
have the form of a (suitably generalized) Chern-Simons (CS) theory in
$2+1$-dimensions \cite{wijo,mose,csref}. This powerful approach
has been very successful in the description of general Abelian FQH states.
However, for the case of non-Abelian FQH states, such as the Pfaffian
state and its generalizations, writing down such a CS theory has proven to be
rather more subtle
\cite{wen-projection,cspf2,cspf1,laplata,frohlich-coset,schoutens}.
One of the outstanding questions
is the connection bewteen the well-defined CFT notion of
non-Abelian statistics and its $2+1$-dimensional
counterpart.

In this paper we construct a class of CS theories that
is appropriate for describing the bulk excitations of
the Pfaffian states and their parafermionic generalization.
The theories that we will discuss here contain
two Abelian fields. One field describes the charge quantum numbers
of the quasiparticles, while the other field is associated with a set of
additional quantum numbers decribing the ``neutral sector''. 
The charged sector is
given by the familiar construction that yields,
for example, the usual Laughlin states \cite{wen}.
The novelty of our approach relies in the choice
of theory describing the neutral sector. It is
given by an Abelian CS theory with a gauge
group with {\it disconnected} components,
specifically $O(2)$. These theories are appropriate for
describing the $2+1$-dimensional analogs of
the relevant Abelian orbifold CFTs \cite{mose}.
Since the Pfaffian models can be also conveniently
realized as orbifold CFTs \cite{cgt}, our approach yields
a simple way of dealing with the bulk theories
corresponding to the Pfaffian states. Moreover, we could also include in our
approach some
generalizations of the Pfaffain states, most
notably some of the Parafermionic states.

One advantage of this formulation is that, although it makes no explicit
assumption
on the nature of the microscopic correlations responsible for the non-Abelian
statistics, it gives new insights on the meaning of the non-Abelian statistics
in these
systems. Specifically, for a $O(2)=SO(2)\times{\bf Z}_2$
CS theory at level $k$, a symmetry ${\bf Z}_{2k}$
characterizes the different disconnected components.
One imposes the condition of charge neutrality in
this sector by moding out this symmetry by the ${\bf Z}_2$
symmetry of charge conjugation, resulting in a
${\bf Z}_k$ symmetry that characterizes the
neutral sector of the complete theory.
This construction defines an {\it Abelian orbifold}
in the context of CS gauge theories \cite{mose}.
In this context, the level $k$ is interpreted as the number
of intermediate channels in the fusion rules
of the CFT description \cite{cft}.
We will discuss below that the level of this theory may in fact be
related naturally with the physics of pairing (and its generalizations).
In the CS version of the theory, we find that it is necessary
to introduce a set of Wilson loop
operators with fractional magnetic flux
in order to insure the completeness of the theory.
For these operators to make sense, one is
led to consider a multiple $k$-covering
of the $2+1$-dimensional space-time
manifold \cite{diwi}.

Equivalently, one could also
define a special class of Wilson loops
with a specific regularization
that involves the twisting of the framing
field \cite{wijo} by $k$ units.
It is this multivaluedness of some of the
Wilson loop operators that is the root of the notion
of non-Abelian statistics, that follows from
the CFT definition: more than one intermediate
channel is involved in the fusion rules. In
the CS theory, the fusion rules follow simply
from charge conservation, modulo the multivalued
character of some of the operators.
The physical mechanism of non-Abelian statistics
in this context arises, therefore, from considering
the identification of
several copies of identical (Abelian) theories,
by some projection mechanism \cite{cspf2}.

In section 2 we review the Abelian CS theory and its
connection to CFT, stressing the points that are crucial
for generalizing the theory to $O(2)$ gauge groups.
In section 3 we discuss at some length the CS theory that
corresponds to orbifold CFTs. More specifically, we
discuss the $2+1$ dimensional definition of the $N$
toroidal models \cite{dvv,dvvv}, and their
${\bf Z}_2$ orbifolds. These models turn out to be
relevant for the Pfaffian systems.
In section 4 we consider specific applications to the
quantum Hall effect, such as the Pfaffian and Parafermionic
states, and their CS formulation.
Our conclusions are summarized in section 5.

\section{Abelian Chern-Simons Theories}

To describe the charged sector we consider an Abelian Chern-Simons
field theory, defined by the action
\beq
S\ =\ {\frac{k}{4\pi}}\ \int_{\cal M}\ d^3x\ \varepsilon^{\mu\nu\rho}
a_{\mu}(x) \partial_{\nu} a_{\rho}(x)\ ,
\label{csac}
\eeq
where $a_{\mu}(x)$ is the Chern-Simons field, $\cal M$ is the
space-time manifold on which coordinates $x$ are defined,
and the coupling constant $k$ is a non-negative integer,
which is referred to as the level.
Having in mind physical
applications, we consider ${\cal M}=\Sigma \times S^1$, where
$\Sigma$ is a two-dimensional spatial manifold (a Riemann
surface) and $S^1$ is the compact time coordinate.
For our applications, $\Sigma$ is taken to be a manifold
without curvature, which could have a boundary (e.\ g.\ , a disk or
an annulus) or not (e.\ g.\ , a torus).
The gauge field $a_{\mu}(x)$ represents a set of matter currents,
as is the case in all effective theories of the 
FQHE \cite{wen,frohlich}
Consequently, the conserved hydrodynamic current
\beq
J^{\mu} (x)\ =\ \frac{k}{2\pi}\ \varepsilon^{\mu\nu\rho}\
\partial_{\nu} a_{\rho} (x)\ .
\label{curr}
\eeq
couples minimally to an externally applied electromagnetic field.

The topological theory defined by (\ref{csac}) has
global observables, given by the Wilson loops
\beq
W_n[C]\ =\ \exp\left(i\ n\ \Phi[C]\ \right)\ ,\ \Phi[C]\ =\
\oint_C\ dx^\mu a_{\mu}(x)\ ,
\label{wilo}
\eeq
where $C$ is a closed path in $\cal M$, $\Phi[C]$ the magnetic
flux subtended by any surface based on $C$ and $n$ labels the 
$U(1)$ representation.
We consider that $n$ takes all possible integer values, which
means that we can measure and distinguish the value
of different charges by coupling the system to an external source.
Therefore, the Hilbert space of the theory on a closed spatial manifold
$\Sigma$ is constructed out from a collection of `bulk' states
$|\ n\ \rangle$, which are labelled by the holonomy $n$.
The vacuum expectation values of the Wilson 
operators furnish a representation of the
one-dimensional lattice of fluxes (or magnetic charges):
\beq
\langle\ W_n[C]\ \rangle\ \cdot \langle\ W_m[C]\ \rangle\
=\ \langle\ W_{n+m}[C]\ \rangle\ .
\label{qcon}
\eeq

The spectrum of the theory is obtained \cite{wen-topological}
by minimally coupling the current
(\ref{curr}) to an external electromagnetic field, and integrating out
the field $a_{\mu}(x)$, yielding
\beq
Q_n\ =\ \frac{n}{k} \qquad , \qquad
{\frac{\theta_{n}}{\pi}}\ =\ {\frac{n^2}{k}}\ ,
\label{spec}
\eeq
where $n$ is an integer and $Q_n$ and $\theta-{n} /{\pi}$ are the
electric charges and quantum statistics of the
quasi-particle excitation corresponding to the state
$|\ n\ \rangle$.

The connection of the above theory to the corresponding
$1+1$-dimensional CFT is established along the lines of the classic
work of Witten \cite{wijo}. Namely, we consider a (compact) spacetime manifold
${\cal M}=D \times S^1 $, where $D$ is the spatial disk and $S^1$ represents
the (compact) time coordinate $t$. The
presence of
the boundary breaks general covariance in $2+1$ dimensions.
Therefore, the topological degrees of freedom of the Chern-Simons theory
become dynamical variables of the corresponding $\u1$ CFT describing
the $1+1$-dimensional chiral bosonic field
$\varphi (\theta + i v t)$, with $R,\theta$ the radius and angular
coordinate of the disk and $v$ the velocity of the edge excitations.
This free theory has a central charge $c=1$ and Lagrangian density
given by
\beq
{\cal L} = {\frac{1}{4\pi}}\ \left( \partial_t \varphi
-v \partial_x \varphi \right) \partial_x \varphi\ ,
\eeq
where $x=R\theta$ and we have arbitrarily chosen one chirality.
In the following, it will be useful to define a holomorphic
coordinate $z=\exp(-i\theta -v t/R)$ on the entire complex
plane.
Arbitrary correlators are evaluated using the basic two-point function
normalized as
\beq
\langle\ \varphi (z) \varphi (w) \rangle\ =\ -\ln (z-w)\ .
\eeq
The observables are given by vertex operators
\beq
V_{\ell}(z)\ =\ {\bf :} \exp\left(i\ \frac{\ell}{\sqrt{k}}\
\varphi (z) \right) {\bf :}\ \qquad \ell\ \in {\bf Z}\ .
\label{vexop}
\eeq
The $\u1$ symmetry of the theory is manifest through the
presence of the current
\beq
J(z)\ =\ - \frac{1}{2\pi}\ \partial_z \varphi(z)\ ,
\label{cutd}
\eeq
which is the `boundary' analog of the `bulk' expression
(\ref{curr}).
The spectrum of charges and conformal dimensions spanned by
these operators is given by $Q_{\ell}={\ell}/k$ and
$h_{\ell}={\ell}^2/2k$, respectively.

Note that it is important to consider the topology of
the Abelian gauge group manifold to discuss the class of observables:
if the gauge group is ${\bf R}$, the theory is not a rational
CFT (RCFT) \cite{cft}, because the number
of primary fields, which in this case are in one-to-one
correspondence with the vertex operators (\ref{vexop}), is
infinite (and parametrized by $\ell$). One can equivalently
define a set of conformal highest weight states
$|\ \ell\ \rangle\ =\ V_{\ell} (0)\ |\ 0\ \rangle$, where
$ |\ 0\ \rangle$ is the standard $SL(2,{\bf C})$ vacuum of
CFT \cite{cft}.
Physically, $|\ \ell\ \rangle $ can be put in correspondence with the
CS state associated to the Wilson operator $W_{\ell}[C]$ that
carries magnetic flux and therefore can be thought of
as adding $\ell$ `vortices' to the vacuum.
On the other hand, if the gauge group manifold is compact, such as
${\bf R}/{\bf Z}$, then the field $\varphi (z)$ is compactified
on a circle of radius $r$, that is
$\varphi (z) \equiv \varphi (z) + 2\pi r $; the value of $r$ is
obtained by demanding that all vertex operators $V_{\ell} (z)$
are well-defined under this shift, which implies $r=\sqrt{k}$.
If $k$ is an integer, the spectrum of the theory is finite:
$Q_{n+k} = Q_n + 1$, $h_{n+k} = h_n + k/2 + n $. This means
that there are only $k$ independent primary fields of
the form (\ref{vexop}) and the
theory possesses a symmetry ${\bf Z}_k$. It is, therefore,
a RCFT. Note that if one imposes the condition that the quantum
statistics $\theta/\pi$ of the corresponding quasi-particles
in $2+1$ dimensions (which is defined up to an even integer)
is invariant under this identification, then $k$ must be even.
In physical terms, this means that there are $k$ different independent
quasi-particles, that add up to a $k$ composite one that can
be created or removed from the system without changing the
vacuum.

Hence, the connection between CFT and
Abelian Chern-Simons field theory
can be summarized as follows \cite{wijo}\cite{mose}:
\begin{itemize}
\item
The coefficient $k$ in the CS action and the square of the compactification
satisfy $k=r^2$;
\item
The Wilson loops correspond to vertex operators;
\item
Charges are proportional (equal with a convenient definition), i.e. $\ell=n$;
\item
Conformal dimensions and quantum statistics of the quasi-particle
excitations are related by $2h=\theta/\pi$;
\end{itemize}
Conformal dimensions can be read off from CS data by considering a thickened
closed Wilson loop, defined by a {\it framing} of the original curve $C$.
A framing is a vector field normal to the curve, which is needed in order to
properly define the holonomies in the quantum theory \cite{wijo}.
The writhing number
of this field along the curve is know as the self-linking number,
and could be equivalently defined as the linking number between the
original loop $C$ and the one obtained from it by the displacement
action of the framing. Note that the framing should be regraded as
a regulator
field, in order to make sense of otherwise ill-defined operators
in the theory. It could also be regraded as a generalization
of the familiar `point-splitting' regularization in quantum field theory.

According to \cite{wijo} , the conformal dimension of a primary field
is obtained by shifting the framing in $t$ units, so that
\beq
\langle\ W_n[C]\ \rangle\ \longrightarrow\ \exp\left(2\pi i t h\right)\
\cdot\ \langle\ W_n[C]\ \rangle\  .
\label{tdcd}
\eeq
Therefore, one defines the conformal dimension of the Wilson loop
as $h=\alpha/2\pi$, where $\alpha$ is the $2+1$-dimensional holonomy
in (\ref{tdcd}), with $t=1$.
For the Abelian theory (\ref{csac}), it is given by
$h=n^2/2k$. Note that underlying this definition is the
important fact that after going around the loop $C$ once,
the framing vector field returns to its original value
(as implied by $t=1$). Later on, we will modify this later
condition in order to allow for more general Wilson loop
operators.

\section{Non-Abelian Generalizations}

Another class of interesting theories arises when one establishes
an equivalence relation among the Wilson loops charges $n$. One may
encounter a situation in which not all the values of $n$ are actually
physically different. The simplest case is to consider that $n$
and $m$ describe equivalent Wilson loops provided that $m=n$ mod $2N$
where $N$ is a given natural number. This equivalence relation
introduces a ${\bf Z}_{2N} $ symmetry among the observables
(\ref{wilo}) and gives
rise to a finite number ($2N$) of observable Wilson loop charges.
The values of the holonomies of Wilson unknotted loops $C$
(assuming a canonical framing in $\cal M$) have to be given, therefore,
by the ${2N}-th$ roots of unity:
\beq
\langle\ W_n[C]\ \rangle\ =\ \exp\left(\ 2\pi i\ {\frac{n}{2N}}\ \right)
\ ,\ n\ =\ 1,\dots,2N\ .
\label{holcan}
\eeq
The action for this CS theory is given by (\ref{csac})
with $k=2N$. In principle, this theory possesess an infinite
number of observables, spanned by Wilson loops with
charges $Q=n/2N$ and statistics $\theta/\pi=n^2/2N$.
However, the equivalence relation is
not simply a matter of definition: only a finite
number of bound states are assumed to be observable,
in the sense that not all possible values
of the charge $n$ are observable. In particular,
the vacuum should be identified with the state
$|\ 2N\ \rangle$.
So, even though we can couple the system to external
sources by means of the current (\ref{curr}), one could
only distinguish their ${\bf Z}_{2N}$ structure.
Furthermore, one can consider a additional ${\bf Z}_2$ symmetry
among the Wilson
loop operators, namely charge conjugation. It amounts to
sending $W_n[C] \to W^*_n[C]$,
or, equivalently, $\Phi[C] \to -\Phi[C]$. This is also equivalent
to reversing the orientation of Wilson loops.
Note also that the action of this symmetry on the Wilson
loops is equivalent to the map $n \to (2N -n)$.
Finally, we remark that if the CS theory is meant to describe
neutrally charged quasi-particles, electric charges should not
be observable, but quasi-particles could still carry
magnetic flux, which implies a breakdown of Gauss' Law.

The corresponding $1+1$-dimensional theories to the proposed
CS theories with ${\bf Z}_{2N}$ symmetry are the $c=1$ gaussian
${\bf Z}_{2N}$ models, the so-called $N$ {\it toroidal models}
\cite{dvv}\cite{dvvv}.
These are rational CFTs compactified on a circle of
radius $r=\sqrt{2p/p'}$ with $p$ and $p'$ coprime integers
such that $N=p p'$. Under the duality $r \to 2/r$, which
makes sense only after a corresponding isomorphic antichiral
theory is glued to the holomorphic one we are discussing,
the roles of $p$ and $p'$ are exchanged. Without loss
of generality, we therefore make here the choice $p'=1$ and
$p=N$, which implies that $r \geq \sqrt{2}$.
The $N$-toroidal model possesess only $2N$ primary fields,
given by the vertex operators (\ref{vexop}), with $n=\ell=1,\dots,2N$.
Note that since $2N$ is even,
there is a further ${\bf Z}_2$ symmetry which could be
eliminated from the spectrum.
When this is done, it gives rise to an {\it Abelian}
${\bf Z}_2$ {\it orbifold} CFT. The ${\bf Z}_2$ symmetry action on
the chiral field is $\varphi  \to - \varphi$.
Note that enforcing this symmetry automatically describes theories
that are electrically neutral, as can be seen recalling the
expression of the current (\ref{cutd}).
This mechanism is the origin of the non-Abelian
statistics, as will be discussed below.

Let us discuss in greater detail these CFTs.
The family of vertex operators consistent
with the ${\bf Z}_{2N}$ symmetry of the $N$-toroidal
model is given by
\beq
V_{\ell}(z) = {\bf :}\exp\left( i\ {\frac{\ell}{\sqrt{2N}}}\
\varphi(z)\right) {\bf :}\ ,\ \ell\ =\ 1,2,\dots,2N\ ,
\label{veopn}
\eeq
with charges $Q=\ell/2N$ and conformal dimensions $h=\ell^2/4N$.
The corresponding ${\bf Z}_{2N}$ symmetry is given by the
equivalence relation $|\ \ell\ \rangle\ \equiv |\ \ell\ +\ 2N\ \rangle$,
and the compactification radius is $r=\sqrt{2N}$
Under the operation $\varphi \to -\varphi$, $\ell \to 2N - \ell$,
some vertex operators remain fixed and the
rest exchange among themselves. The remaining theory still
possesess a ${\bf Z}_N \simeq {\bf Z}_{2N}/{\bf Z}_2 $ symmetry.
The CFT obtained by moding out the ${\bf Z}_2$ symmetry
in the toroidal models is a chiral orbifold \cite{dvvv}.
These are further CFTs with $c=1$ and the same
compactification radius as the starting gaussian models, but
that only retain the sectors of the toroidal Hilbert space
which are consistent
with the ${\bf Z}_2$ discrete symmetry.
Moreover,
the introduction of additional primary fields
is required
for consistency of the theory. In CFT these consistency
conditions are the fusion rules and the modular invariance
of the partition function when $\Sigma$ is a annulus \cite{wijo}
\cite{mose}.
Let us summarize here the results of \cite{dvvv} for a
holomorphic model: the additional primary fields
generate the `twisted sector', which contains fields
with anti-periodic boundary conditions
\footnote{
This statement applies to a spatial manifold
with the topology of a cylinder. On the complex
plane, into which it is mapped by a standard conformal
transformation, the fields have periodic boundary conditions.
This is a well-know phenomenon \cite{cft}.}
(for either spatial or - through modular invariance -
time-like closed curves), and is independent
of the value of $r$.
The additional operators add sectors with half-integer
charges, and create a new ground state out of the vacuum
for fields allowed by the enlargement of acceptable
boundary conditions. This new sector is known as the
Ramond sector in the string theory literature \cite{cft},
as opposed to the Neveu-Schwarz sector, which pertains
fields with periodic boundary conditions.
These operators are generically know as `disorder operators'
because of the role they play in the Ising model, and are
given by
\beq
\sigma(z)\ =\  {\bf :}\exp\left( i\ {\frac{1}{2\sqrt{2}}}\
\varphi (z)\right)
{\bf :}\ ,\
\tau (z)\ =\ {\bf :}\exp\left( i\ {\frac{3}{2\sqrt{2}}}\
\varphi(z)\right)
{\bf :}\ ,
\label{dista}
\eeq
of conformal dimensions $h=1/16$ and $h=9/16$, respectively.
These operators are non-local in the original toroidal model,
but become local in the orbifold.
One way of viewing them is as the only two allowed
operators with half-integer charges in the smallest
($N=1$) toroidal model, with $r=\sqrt{2}$. This is because
conformal dimensions of (holomorphic) primary fields should
be less than 1 in the $N=1$ toroidal model, which can
in turn can be rephrased as the statement that (\ref{dista})
are the only two relevant operators (in the sense of the
renormalization group) that could be added
with half-integer charges \cite{dvvv}. As stated before,
these operators generate the twisted sector of the ${\bf Z}_2$
Abelian orbifolds for all $N$.

What are the corresponding $2+1$-dimensional disorder operators?
To answer this question one should be able to make sense of the
analogs of the disorder operators (\ref{dista})
that describe excitations with half-integer
flux values.
Consider a Wilson loop (\ref{wilo}) for $n=1/2$:
\beq
W_{1/2}[C]\ =\ \exp\left(i\ \frac{1}{2}\ \Phi[C]\ \right)\ ,
\label{hawil}
\eeq
in a CS theory (\ref{csac}) with $k=2N$, corresponding to
the $N$-toroidal CFT. Consider now the action of twisting
$t$ units the framing field when traversing the closed loop
$C$ in (\ref{hawil}) once. The holonomy of the vacuum expectation
value of $W_{1/2}$ is shifted according to (\ref{tdcd}), with
\beq
\alpha= 2\pi\cdot t \cdot \frac{1}{16N}\ .
\eeq
We notice that the desired conformal dimension $h=1/16$
is obtained from the previously stated general
expression $h = \alpha/2\pi$, provided that $t=N$,
for all $N$.
A similar reasoning yields a definition of the operator
$W_{3/2}$, with $h=9/16$.
The meaning of the newly defined Wilson loops is as follows:
the framing field provides a way of giving sense to the
$2+1$-dimensional analog of a multi-covering of the complex
plane \cite{ze}.
Fields with no branch-cuts in the complex plane
are lifted to Wilson loops with $t=1$, whereas fields with
branch cuts of order $N$ are lifted to Wilson loops 
that are defined with $N$ twists
when traversing the loop once.
This is the first time in this work we see the appearance
of $N$ copies of a given underlying structure, but this
idea will arise again in discussing the physical applications
of this class of CS theories.
One may also ask what is the vacuum holonomy of the operator
$W_{1/2}$. We argue that is is given by (\ref{holcan}) with
$n=1$, because it is the minimum amount of charge (flux)
that exists in the CS theory. However, multivaluedness
of the Wilson operator implies that this charge
should be considered as evenly distributed over the
$N$ different coverings of the loop, as represented
by the framing field.

One can gain further support for the correctness of the above
definition of the CS theory by considering the consequences 
of modular covariance of the orbifold CFT, as viewed 
in the CS formulation.
Consider the CS theory defined on the spacetime manifold 
${\cal M}=S^1 \times S^1 \times I$,
where the space manifold is given by an 
annulus $\Sigma= S^1 \times I$, with
inner (outer) radius $R_R$ ($R_L$), respectively and $I=[R_L, R_R]$. 
Completeness of the theory requires that 
the partition function, which
includes both chiral (say, for
the outer edge excitations) and anti-chiral sectors (inner
edge), should be taken to be modular invariant 
under the congruence subgroup $\Gamma(2N)$
of the modular group $PSL(2, {\bf C})$ \cite{dvv} \cite{cz},
instead of the full modular group, in order to
accomodate for the disorder operators.
Given the general
covariance of the CS theory, one may exchange space and
time coordinates, and obtain a torus
$S^1 \times S^1$, which has been considered in \cite{wijo}.
On this torus with given modulus $\tau$
the generators of the modular group $T$ and $S$ 
act as $T:\ \tau\ \to\ \tau +1 $ and
$S:\ \tau \to\ -1/\tau$. $T$ can be thought of as the
generator of closed paths in $\Sigma$ for loops at fixed
$2+1$-dimensional time; as such, it probes the boundary
conditions of the conformal fields.
An explicit realization of the $T$ and $S$ operators in terms
of the CS data is given by the Verlinde operators
\cite{ver}.
Note, however, that the relevant subgroup in our discussion is
$\Gamma(2N)$ which is generated by $T^{2N}$ and $ST^{2N}S$. 
This requirement is also the same to impose on the canonical 
quantization
of the CS theory \cite{wijo}\cite{djt}, in which a constant time 
surface is considered.
In this setting, the action of $T$ can be associated to
the framing field. Therefore, formulating a theory with 
disorder operators in this framework also amounts to 
take again $T^{2N}$ as
a generator, rather than $T^2$ (which would be the standard
requirement for a theory with fermionic excitations).

What are the resultant CS theories corresponding to the
${\bf Z}_2$ orbifolds of the ${\bf Z}_{2N}$ gaussian theories?
The answer to this question has been discussed in Refs. \cite{mose}
and \cite{diwi}:
these correspond to theories with a gauge group $G=O(2)$ and level
$k=2N$. This can be understood in view of the previous discussion:
denoting by $G_k$ the level $k$ CS theory (\ref{csac})
with $G\ =\ U(1)\ \simeq\ SO(2)$, one has that
$O(2)_k\ \simeq\ \left(\ SO(2)\ \times\ {\bf Z}_2\ \right)_k\
\simeq\ U(1)_k\  \times\ {\bf Z}_{2k}$. After imposing
charge conjugation invariance (i.e., the orbifolding
procedure), one is left with a discrete ${\bf Z}_k$ symmetry.
From now on, we will refer to the $2+1$-dimensional
theory corresponding to the ${\bf Z}_2$ orbifold
simply as a $O(2)_N$ CS theory. Note, however, that in doing
so one implicitly assumes the previous definition of the CS disorder 
operators.

Finally, we would like to remark that one could also view these
models as Topological
Spin Theories \cite{diwi}. These are theories in which the
space-time manifold $\cal M$ admits an spin structure. The framing
fields allow one to think of these manifolds as multi-covered
standard ones, in the sense implied by the definition
of fractionally charged Wilson loops. These manifolds were
introduced as a natural framework in
which $2+1$-dimensional topological field theories can be
defined when one considers a gauge group $G$ wich is not connected.

We now would like to discuss some examples in more detail:

\subsection{$N=1$}

The first example we would like to discuss is the case $N=1$.
Consider first the CFT $N=1$ toroidal theory: it is
a gaussian theory ($c=1$) with compactification radius $r=\sqrt{2}$.
Before the orbifolding procedure is applied, consider
the three primary fields: $J^3(z)= i\partial \varphi (z)$,
$J^{\pm}(z)= {\bf :}\exp( \pm i \sqrt{2} \varphi(z) ){\bf :}$ of
conformal weight $1$, which satisfy an $\su2$ current algebra.
The theory possesess two vertex operators
of the form (\ref{veopn}), namely $V_1 (z) = \exp
(\frac{i}{\sqrt{2}} \varphi(z))$ and $V_2(z)= J^+(z)$ of
charges 1/2 and 1, and conformal dimensions 1/4 and 1,
respectively. They correspond to the two representations
of the affine $\su2$ algebra, namely the identity $[1]=[V_2]$
and the spinor $[\Psi]=[V_1]$, with ${\bf Z}_2$ fusion algebra
$\Psi \times \Psi = 1$. The ${\bf Z}_2$ orbifold of this
theory is known to be equivalent to the $N=4$ toroidal
model \cite{dvvv}. A counting of representations
in the theory is consistent with the ${\bf Z}_8$ symmetry
of the resulting theory: there are four representations
arising from the twisted sectors (two for each of
the operators (\ref{dista})), and four arising from
the untwisted sector ($k=2$ times for each $V_i$, see
\cite{dvvv}).

The CS theory of the toroidal model consists of two clases of
Wilson loops, namely $W_1[C]$ and $W_2[C]$ with holonomies
$(-1)$ and 1, respectively. These operators correspond to the
CFT vertex operators $V_1$ and $V_2$, respectively.
The holonomies are real, and charge conjugation does not
change their values, so both states $|\ 0\ \rangle$ and
$|\ 1\ \rangle$ are fixed points of the ${\bf Z}_2$
transformation. Therefore, both states are retained
in the `orbifold' theory.
This result is consistent with the expectation
that there is no discrete symmetry surviving the ${\bf Z}_2$
projection. The Abelian fusion rule of the toroidal
CFT is
reproduced by charge (flux) addition of the Wilson loops
holonomies Eq. (\ref{qcon}), i.e,
$\langle W_1 \rangle \cdot \langle W_1 \rangle
= \langle W_2 \rangle = 1$.
In addition, both disorder operators $W_{1/2}$ and
$W_{3/2}$ should be added, but in this case
one should require $t=1$, as for the rest of the
Wilson loops. Therefore, no special condition
distinguishes these operators in this case.
On the other hand, consider the CS theory
with $k=8$. The spectrum of Wilson loop charges and
conformal dimensions is given by (\ref{spec}). We
see that the corresponding Wilson loop operators
$W_n$, $n=1,\dots,4$ correspond one-to one to the
previously discussed set. This indicates
that the postulated correspondence between $2+1$ and
$1+1$ dimensional theories is consistent with the
corresponding orbifolding procedures in this example.

\subsection{$N=2$}

The second example, for $N=2$, is interesting
because it appears in the neutral sector in Pfaffian
states of the quantum Hall effect. We first consider
the $N=2$ toroidal model CFT, with compactification
radius $r=2$.
The ${\bf Z}_4$ structure of the theory is given by the
four vertex operators $V_{\ell} (z) =
{\bf :}\exp\left( i\ {\frac{\ell}{2}}\
\varphi(z)\right) {\bf :}$, with $\ell = 1,2,3,4$, of carges
$Q=1/2,1,3/2,2$ and conformal dimensions $h=1/8,1/2,9/8,2$,
respectively.
The ${\bf Z}_2$ orbifold theory yields
two identical copies of the Majorana
fermion CFT (the Ising model at the critical point),
such that $c=1=1/2+1/2$.
A description of this theory in terms of a chiral
boson has been discussed in \cite{dvv}\cite{kiri}.

The CS theory of the $N=2$ toroidal model has,
correspondingly, four classes of Wilson
loops, $W_\ell [C] = \exp\left( 2\pi i \ell \Phi[C] \right)$,
with holonomies $\exp{\left(i\pi\ell/2\right)}$,
${\ell}=1,2,3,4$.
Under charge conjugation, $W_1 \to W_3$ with $W_2$ and
$W_4$ fixed. Orbifolding amounts to rendering the
theory electrically neutral and introducing the
`disorder' operators $W_{1/2}$ and $W_{3/2}$
defined in (\ref{hawil}),
of conformal dimension $1/16$ and $9/16$, respectively.
Now, for these operators to make sense, we need to
define them by requiring that the framing field
undergoes a twist of $t=2$ units when traversing
the loop $C$ once, as discussed above.
This is the first CS example in which we encounter this
new feature.
The other independent Wilson loop operators are
$W_2 = \psi$ and $W_4 \equiv W_0 = 1$,
of conformal dimensions $1/2$ and $2 \equiv 0$,
which is the operator content of the Ising model
($\psi$ is the Majorana field).
A minimal set of operators under the Abelian
composition law of Wilson loops (\ref{qcon}) is
given by $W_{1/2}$, $W_2$ and $W_4$.
This is because $W_2$ and $W_4$ are invariant
under charge conjugation, and $W_{1/2}$ is
mapped onto $W_{3/2}$, after taking into
account that the actual charges of both
operators are $1$ and $3$, respectively,
as argued before.
Note that $W_2$ and $W_4$ furnish a representation of
the surviving ${\bf Z}_2$ symmetry.

Furthermore, one realizes that there are actually two copies
of the same theory: although the holonomies
(\ref{holcan}) of $W_2$ and $W_4$ are real, the holonomy of
$W_{1/2}$ is $\pm i$, after considering the double
covering of $\cal M$.
This sign ambiguity means that there actually two
equivalent ways of realizing this theory.
More precisely, in the CFT formulation of the Ising
model \cite{cft}, one decomposes the Weyl fermion of a
parent $c=1$ CFT into two identical but distinct
Majorana fermions, each described by a $c=1/2$ CFT.
Next, one introduces a {\it projection}, i.\ e.\ , a reality
condition that retains only one of the two Majorana
fields as a legitimate degree of freedom. Both possible
choices yield equivalent results due to the original
symmetrical decomposition. Therefore, the $c=1 \to
c=1/2$ projection in CFT is a mechanism of {\it halving}
the theory. Although performing a general projection
in CFT calls for a coset construction \cite{cft}, in the
particular case we are considering this procedure is
not necessary. Correspondingly, we apply an analog
approach at the level of the CS theory: the doubling of degrees of
freedom analog to the one occuring in the
$c=1=1/2+1/2$ CFT is paralelled by the existence of
two equivalent Wilson loop operators $W_{1/2}$,
which could be distinguished by their canonical
holonomies $\pm i$. In view of the previous
discussion, we adopt as a natural projection
condition the picking of
a sign determination for the operator $W_{1/2}$,
and claim that this realizes in $2+1$ dimensions
the corresponding projection $c=1 \to c=1/2$ in
$1+1$ dimensions.

Moreover, one can also verify the crucial property of the
reproducibility of the CFT
fusion rules. These are given by charge addition of
the corresponding Wilson loops, modulo the issues
concerning the multivaluedness of the $W_{1/2}$ operator.
Therefore, the product
$W_2 \cdot W_2 = W_4$ verifies the corresponding
CFT rule $\Psi \times \Psi = 1$.
The product $W_{1/2} \cdot W_2 = W_3 \simeq W_{3/2} = W_{1/2}$
verifies the fusion rule $\sigma \cdot \psi = \sigma$,
after taking into account the correct charge assignments,
and the identification $W_3 \simeq W_{3/2}$ actually means
that the operator in the rhs is multivalued, whereas
the one with the same charge in the lhs is not; in
the same sense, one has $W_1 \simeq W_{1/2}$.
The product $W_{1/2} \cdot W_{1/2}$ is the most
interesting of all. Due to the double covering
associated to each operator, one should consider
actually all possible choices of charge addition.
Charges should also be provided with a second label
indicating in which covering they are considered.
This label could be a sign, since the group
involved is ${\bf Z}_2$.
This makes the charges actually behave like the third
component of an $SU(2)$ multiplet.
Is in now clear that the possible outcome now
for the addition of charges is $0$ or $2$,
given that each indivual charge is $1$.
We indicate this by writing
$ W_{1/2} \cdot W_{1/2} = W_0 + W_2$,
which reproduces the CFT fusion rule
$\sigma \cdot \sigma = 1 + \Psi$.

\subsection{$N \geq 3$}

The ${\bf Z}_2$ orbifolds of the $N$ toroidal models
for other values of $N$
can be analyzed along the same lines. We remark here
that for $N=3$, the ${\bf Z}_4$ parafermionic model
is obtained \cite{dvvv}. This model is of relevance
for further Quantum Hall universality classes known
generically as {\it Parafermionic States}
\cite{rere}. However, more general parafermionic
states are not obtained by considering further
${\bf Z}_2$ orbifolds of the $N$ toroidal models.
This can be understood from the fact that the
$K$ parafermion models are described by CFTs
with central charge $c=2(K-1)/(K+2)$, which is
larger than $1$ for $K > 4$, and therefore
beyond the scope of the systems considered
here.
Further examples of the models considered
here are also known: for $N=4$ one obtains the
four-state
Potts model, and for $N=6$ a discrete superconformal
model at $c=1$ \cite{dvvv}.

\section{Applications to the Quantum Hall Effect}

For applications to the quantum Hall effect, we consider
a direct product of two Abelian CS theories, corresponding
to the charged and neutral excitations.
In principle, one has several choices among the different
possibilities we have discussed above. A consistency
principle is therefore required.
A usual constraint is to require the presence of an
excitation with the quantum numbers of the electron
in the spectrum of charges and quantum statistics of
the combined theory. However it is not clear what additional
conditions (if any) are required in general to determine
uniquely the physically acceptable theories.

\subsection{The Pfaffian states}

Let us now consider the $q$-Pfaffian states \cite{more}.
For $q$ even (odd), these models are
known as fermionic (bosonic) Pfaffian theories.
These theories can be constructed from two Abelian CS
theories: one for the charged $(+)$ and another for
the neutral $(0)$ sectors, respectively.
The charged Abelian CS field is defined with (positive) integer
level $k=q$. The physical requirement on the spectra
of these theories is that there exists a particle
with unit electric charge and odd quantum statistics
(electron or hole) for $q$ even. A similar
statement, with even quantum statistics, is
needed for $q$ odd. In other words,
the `electron' excitation has unit charge and
fermionic ($q$ even) or bosonic ($q$ odd)
statistics.
Clearly these conditions cannot be fulfilled by
the charged sector alone, with quasiparticle spectrum
given by (\ref{spec}), namely
$Q=n/q$ and $\theta^+ /\pi = n^2/q$.
One therefore considers a
suitable choice for the neutral sector. Note, however, that
when coupling the combined system to an external electromagnetic
probe, only the charged sector contributes to the Hall
conduction, yielding the value $\nu =1/q$ for the
filling fraction.

A natural CFT description of the neutral sector
of these models is to consider the ${\bf Z}_2$ orbifold of the
toroidal model with $N=q$ \cite{cgt}.
Given that we now know how to construct the $2+1$
dimensional analog of these orbifolds, we consider
an Abelian CS theory at level $k=2q$ for the
fermionic Pfaffian and level $k=2(q+1)$ for the bosonic
case.
Equivalently, the compactification
radii of charged and neutral sectors,
$r_+$ and $r_0$ respectively, satisfy
either the condition $r^2_0 = 2 r^2_+$ (fermionic Pfaffian)
or $r^2_0 =  2(r^2_+ + 1)$ (bosonic Pfaffian).
From previous discussions, the resulting neutral theory has a
${\bf Z}_q$ symmetry in the fermionic case, and
${\bf Z}_{q+1}$ in the bosonic case.
The spectrum of the theory of the neutral sector is $\theta^0 /\pi =
s^2/2q$, where $s$ is integer (half-integer) in the untwisted (twisted)
sector, respectively, for the fermionic Pfaffian.
For the bosonic case, one has $\theta^0 /\pi = s^2/2(q+1)$ instead.
More explicitly, we consider the total CS theory describing
the Pfaffian systems to have symmetry $U(1)_q \times
O(2)_{2q}$ for the fermionic case and $U(1)_q \times O(2)_{2(q+1)}$
for the bosonic one, where the first (second) factor refers
to the charged (neutral) sector.

The total theory combining both the charged and neutral
sectors considered above now admits electrons in its spectrum.
Consider first the case of the fermionic Pfaffian,
and assume that $q/2$ odd.
Then, by choosing $s=q$ in the neutral sector,
one has that the condition in the charged
sector $Q=1$ enforces $\theta^+/\pi = q$, and
$\theta^0 /\pi = q/2$. Adding the contribution
from both sectors, a fermionic excitation can
be made. If $q=4r$, again one can choose
$s=q/2$ to obtain overall odd statistics
provided $r/2$ is odd. Clearly, this
analysis can be extended to cover all
possible cases along the same lines.
The analysis of the bosonic Pfaffian also follows
from a similar reasoning.
Consider first the case $(q+1)/2$ odd. Then,
it is enough to take the same quantum numbers in the
charged sector as in the fermionic case, and $s=q+1$
in the neutral sector to obtain an excitation
with $Q=1$ and $\theta /\pi = q + (q+1)/2$.
Further cases could be analyzed along the same lines.

For the specific case of the $q=2$
fermionic Pfaffian, with filling fraction $\nu=1/2$,
we have the following results:
the electron operator can be constructed out
of two Wilson loops: $W^{+}_1$, with $Q=1$
and $\theta^+ / \pi = 1$ in the charged sector
and $W^0_2$, with no charge and $\theta^0 /\pi=
1/2$ (Majorana field) in the neutral sector.
We consider the Wilson operators for the charged and
neutral sectors as defined along the same Wilson
loop $C$, since the theory is a direct product of
both sectors.

Another known property of these systems is their
topological order \cite{wen}. For our purposes, it is given
by the number of independent states in the CS theory.
For simplicity, we focus the discussion on the
fermionic case.
We have here $q$ independent states for the
different charge sectors, and three states for
the neutral sector.
If all possible pairing of states between
the charged and neutral theories lead to
a consistent theory,
one would in fact have a total of $3q$ states
\cite{mire}.
Actually, $2q$ of these arise from the
untwisted sector of the neutral theory,
whereas $q$ arise from the twisted
sector.
In CFT the corresponding consistency
condition to establish this pairing is given by
modular invariance \cite{cz}.
In $2+1$ dimensions, modular invariance
is a consequence of general covariance
and gauge invariance \cite{wijo}.
This implies that all possibilities should be
taken into account.

\subsection{Non-Abelian Statistics in the Pfaffian states}

In the following, consider the fermionic Pfaffian for the
sake of simplicity.
In CFT, the quasi-hole operator in the $q$-Pfaffian
states is
\beq
\Psi_{qh}(z)\ =\ \sigma(z)\ \cdot\ {\bf :}
\exp\left( i \frac{1}{\sqrt{q}}\ \varphi^+ (z) \right) {\bf :}\ ,\quad
\sigma(z) = {\bf :} \exp\left( i \frac{1}{2\sqrt{2}}\ \varphi^0 (z)
\right ){\bf :}\ ,
\label{qhop}
\eeq
of charge $Q=1/q$ and conformal dimension $h=1/16 + 1/2q$.
For the case $q=2$ these expressions yield $Q=1/2$ and
$h=5/16$. In eq. (\ref{qhop}) $\varphi^+ (z)$ is the
chiral bosonic field of the charged sector and
$\sigma(z)$ is the chiral disorder operator, of total charge
$0$ and conformal dimension
$1/16$, with $\varphi^+ (z)$ being the chiral bosonic field
in the neutral sector.

The corresponding CS quasi-hole operator is,
therefore, given by
\beq
W_{qh}[C]\ =\ W^+_1 [C]\ \cdot\ W^0_{1/2} [C]\ ,
\eeq
with $W^+_1 [C]$ defined by (\ref{wilo}) in terms
of the charged CS field $a^+_{\mu}(x)$ and
$W^0_{1/2} [C]$ given by (\ref{hawil}) in
terms of the neutral CS field $a^0_{\mu}(x)$.
It is important to have in mind that the definition
of both operators depend on $q$ through the
CS action (\ref{csac}). There is a further,
more important and specific dependence of the operator $W_{1/2}$
on $q$, given that it is assumed to be defined after
$q$ twists
of the framing field when traversing $C$ once.
The statement that the $q$-Pfaffian theories posseses
non-Abelian statistics means in our context that the correlation
function of four quasi-hole operators yields more
than one resulting Wilson loop. This statement
is the analog to the CFT result of several
resulting channels of conformal blocks in the
evaluation of the corresponding CFT correlators.
This latter property is a {\it consequence of the
fusion rules} of the primary fields,
which we have already verified.
Since, by construction, the Wilson loops corresponding
to those primary fields satisfy the CFT fusion rules,
the property of non-Abelian statistics is also
present in the $2+1$-dimensional theory.
Note that in establishing the fusion rules among
Wilson operators, the property of multivaluedness
of the operator $W_{1/2}$ is crucial.
The origin of this multivaluedness can be traced
back to the remaining ${\bf Z}_q$ symmetry in
the neutral sector, and could be ultimately
held responsible for the appearance of non-Abelian
statistics in the Pfaffian models.
A similar analysis could be rephrased for the bosonic case.

We conclude by stating that the $2+1$ dimensional
theory describing the fermionic $q$-Pfaffian
is given by a topological theory of two Abelian
CS fields, with gauge group $U(1)_q \times O(2)_{2q}$
(in the charged and neutral sectors, respectively).
The symmetry of the theory in the neutral sector
is ${\bf Z}_q$ and therefore we consider it as the
responsible for the appearance of non-Abelian
statistics.
Similarly, the bosonic Pfaffian has a
$U(1)_q \times O(2)_{2(q+1)}$ gauge group,
with symmetry ${\bf Z}_{q+1}$ in the neutral sector.

\subsection{The $K=4$ Parafermionic states}

As we discussed above, within the class of states
we consider in this paper, one is able to describe
the cases $K=2$ (Pfaffian) and $K=4$ Parafermion
states. Here we give a brief discussion of these
latter cases.

In general, the parafermionic models are the simplest
CFTs with symmetry within the class of chiral algebras
obtained by
adjoining to the Virasoro algebra higher spin primary
fields that correspond to Casimir operators of a
simply laced Lie group \cite{para}. For the ${\bf Z}_K$
parafermionic models, the Lie group is $SU(K)$.
In the CFT description of the theory, two vertex
operators of conformal spin 3 and 4 should be considered
as primary fields, in the $N=3$ toroidal model.
This yields a set of vertex operators with conformal
dimensions $1/12, 1/3, 3/4$ in the untwisted sector
and $1/16, 9/16$ in the twisted sector.
The corresponding Wilson loop operators in the corresponding CS
theory at level $k=K=3$ are given by $W_n$, $n=1,2,3$
in (\ref{wilo}),
and $W_{1/2}$ and$W_{3/2}$ as defined in section 3,
respectively.
Again, a symmetry ${\bf Z}_3$ survives in the neutral
sector, which yields the notion of non-Abelian statistics
generalizing the case of the Pfaffian.

\section{Conclusions}

In this paper we have considered Chern-Simons theories with two Abelian
fields, one for the charged and another for the neutral
sector. We studied the consequences of considering a
group $O(2)$, with two disconnected components, in the
neutral sector, verifying in some detail that this
theory is the $2+1$-dimensional analog of the
${\bf Z}_2$ orbifolds in Conformal Field Theory

An interesting physical consequence of our study
is that it allows one to think of the non-Abelian
statistics from a seemingly unconventional point of view. Indeed,
here we found that the non-Abelian statistics follows from the
${\bf Z}_q$ structure in the neutral sector for
the $q$-Pfaffian models. This result is interesting
because it brings together seemingly different
types of theories displaying non-Abelian statistics,
such as the Pfaffian and Parafermionic models,
the minimal incompressible models of the hierarchical
Hall states \cite{ctz5}, the Landau-Ginzburg 
approaches \cite{cspf2,cspf1}, and
the coset construction \cite{laplata,frohlich-coset}.
In several of these theories, it has been observed
that the non-Abelian statistics
that characterizes these models also arises form
an underlying ${\bf Z}_m$ symmetry in the neutral
sector, as for the case of the minimal incompressible 
models \cite{cz2}. 
The same structure shows up in
the Landau-Ginzburg and coset constructions.
However, these theories do not exhibit in an obvious manner an orbifold
structure. Likewise, the discrete symmetry associated with the
non-Abelian states in some cases emerges from the pairing (or
clustering) mechanism behind the microscopic physics of these FQH
states.
We expect that further connections along these
lines could be established among these these and other related
theories.

\section{Acknowledgements}

This work was supported in part by the National Science Foundation,
grant number NSF DMR98-17941 at UIUC (EF), by  the Fullbright
Fellowship Foundation (GZ) and by a FOMEC Fellowship (Argentina) (MH).

%
\def\NP{{\it Nucl. Phys.\ }}
\def\PRL{{\it Phys. Rev. Lett.\ }}
\def\PL{{\it Phys. Lett.\ }}
\def\PR{{\it Phys. Rev.\ }}
\def\CMP{{\it Comm. Math. Phys.\ }}
\def\IJMP{{\it Int. J. Mod. Phys.\ }}
\def\MPL{{\it Mod. Phys. Lett.\ }}
\def\RMP{{\it Rev. Mod. Phys.\ }}
\def\AP{{\it Ann. Phys.\ }}

\end{document}